\documentclass[conference]{IEEEtran}

\pdfoutput=1
\usepackage{ifpdf}
\usepackage[compress]{cite}

\usepackage[dvips]{graphicx}
\graphicspath{{figs/}}

\usepackage[cmex10]{amsmath,mathtools}
\usepackage{amssymb}
\usepackage{mathrsfs}
\usepackage{dsfont}
\DeclareMathAlphabet{\mathpzc}{OT1}{pzc}{m}{it}
\DeclareMathAlphabet{\mathcalligra}{T1}{calligra}{m}{n}

\usepackage{algorithmic}

\usepackage{array}
\usepackage{blkarray}

\usepackage{stfloats}

\usepackage{url}

\usepackage[usenames,dvipsnames,svgnames,table]{xcolor}
\usepackage{pifont}
\usepackage{fancybox}
\usepackage{todonotes}
\usepackage{soul}
\ifCLASSOPTIONcompsoc
\usepackage[tight,normalsize,sf,SF]{subfigure}
\else
\usepackage[tight,footnotesize]{subfigure}
\fi
\usepackage{hhline}
\usepackage{multirow}
\usepackage{flushend}

\begin{document}

\title{Deterministic Broadcasting and Random Linear Network Coding in Mobile Ad Hoc Networks}

\author{Nikolaos~Papanikos~\IEEEmembership{Student~Member,~IEEE,}
        and~Evangelos~Papapetrou~\IEEEmembership{Member,~IEEE}
        \\
Department of Computer Science \& Engineering, University of Ioannina, Greece
\\
npapanik, epap@cs.uoi.gr
}

\maketitle

\begin{abstract}
Network coding has been successfully used in the past for efficient broadcasting in wireless multi-hop networks. Two coding approaches are suitable for mobile networks; Random Linear Network Coding (RLNC) and XOR-based coding. In this work, we make the observation that RLNC provides increased resilience to packet losses compared to XOR-based coding. We develop an analytical model that justifies our intuition. However, the model also reveals that combining RLNC with probabilistic forwarding, which is the approach taken in the literature, may significantly impact RLNC's performance.
Therefore, we take the novel approach to combine RLNC with a deterministic broadcasting algorithm in order to prune transmissions. More specifically, we propose a Connected Dominating Set (CDS) based algorithm that works in synergy with RLNC on the ``packet generation level". Since managing packet generations is a key issue in RLNC, we propose a distributed scheme, which is also suitable for mobile environments and does not compromise the coding efficiency. We show that the proposed algorithm outperforms XOR-based as well as RLNC-based schemes even when global knowledge is used for managing packet generations.
\end{abstract}

\IEEEpeerreviewmaketitle


\section{Introduction}

\IEEEPARstart{B}{roadcasting} is a cornerstone of many distributed networking protocols in wireless ad hoc networks. From routing\cite{abolhasan-review} to application layer protocols\cite{Mian2009-servicedisc}, broadcasting is used for distributing and collecting information about the random network. Implementing a lightweight and efficient broadcast mechanism is of paramount importance. Over the last years, network coding \cite{Ahlswede} has emerged as an effective approach to enhance the performance of networking protocols. In this context, several researchers have looked into combining network coding and broadcasting in wireless ad hoc networks\cite{fragouli_rlnc,widmer_rlnc-update,mahmood_arlnccf_icc,hou2008adapcode,cho_DRAGON,yang2011-RCODE-jrn,subramanian2012uflood,CodeB,Directional,kadi2008mpr-nc,DiSC,rahnavard2006CRBCast,vellambi2010FTS}.
Among the proposed algorithms, there exist two major approaches that are generic and can be implemented not only to static but also to mobile networks.
The first combines \emph{Random Linear Network Coding (RLNC)} \cite{PracticalNC} with probabilistic forwarding. More specifically, packets are grouped in the so called ``generations". Encoded packets are produced as random linear combinations of the packets in a generation, based on the theory of finite fields\cite{linear,Ho_randomized}, and then probabilistically forwarded. Receiving enough linear combinations allows the decoding of the original packets. The foundations of this approach have been laid by Fragouli et al.\cite{fragouli_rlnc}.
The key idea is to use RLNC for providing delivery efficiency while probabilistic forwarding alleviates the cost of broadcasting in terms of transmissions.
The second approach works on the concept of ``coding opportunity"\cite{COPE} and encodes packets on a hop-by-hop basis using bitwise XOR (\emph{XOR-based coding}). In contrast to the previous approach, the encoded packets are deterministically forwarded and the coding method is oriented towards reducing transmissions rather than coping with transmission failures. The first and most representative algorithm of this category, proposed by Li et al.\cite{CodeB}, utilizes XOR-based coding while the encoded packets are forwarded with the partial dominant pruning algorithm\cite{PDP}.

In this work, we confirm, through analysis, that RLNC provides increased resilience to packet losses. This feature is critical for coping with the inherently unreliable wireless transmissions. However, in wireless ad hoc networks, the cost of broadcasting, expressed by the number of transmissions, is equally important due to the limited resources. Despite the current trend in the literature, we show that using probabilistic forwarding to suppress transmissions may significantly impair the performance of RLNC. Therefore, we follow the \emph{innovative approach of integrating RLNC into deterministic broadcasting}. Below, we summarize our contributions:
\begin{itemize}
  \item We develop an analytical model (Section~\ref{motivation}) that sheds light on the differences between RLNC and XOR-based coding. Such information is very useful since only sparse empirical data exist in the literature for comparing the two methods. The model confirms that RLNC exhibits increased resilience to transmission errors.
  \item We use the developed model to unveil the potential pitfalls of combining RLNC and probabilistic forwarding.
  \item Following our observations, we turn to deterministic broadcasting, which has never been tested with RLNC. More specifically, the proposed algorithm (Section~\ref{RLDP}) implements CDS (Connected Dominating Set) based forwarding rules ``on the generation level" in order to allow the flow of packet generations over the CDS. The rationale is that the CDS will provide a more systematic pruning of redundant transmissions without impairing the coding efficiency of RLNC.
  \item We address the problem of \emph{generation management}, i.e. the need of nodes to distributively agree in the grouping of packets into generations. Although this is vital for practically implementing RLNC, especially when packets from different sources are ``mixed" into a generation, it is rarely discussed in the literature. We review the pending problems and propose a distributed mechanism that does not compromise the coding efficiency (Section~\ref{generation_management}).
\end{itemize}
\noindent In Section~\ref{related_work}, we discuss the related work. We conclude this work by evaluating the proposed algorithm through simulation (Section~\ref{evaluation}) and summarizing our findings in Section~\ref{conclusion}.

\section{Related Work}\label{related_work}

Several studies have investigated the use of network coding for broadcasting in wireless ad hoc networks. The proposed algorithms can be classified, based on the coding method, into: i) RLNC-based, and ii) XOR-based approaches. RLNC-based algorithms\cite{fragouli_rlnc,widmer_rlnc-update,mahmood_arlnccf_icc,hou2008adapcode,cho_DRAGON,yang2011-RCODE-jrn,subramanian2012uflood} build on the concepts of practical RLNC \cite{PracticalNC} and probabilistically broadcast random linear combinations of packets.
Fragouli et al.\cite{fragouli_rlnc}, extend the probabilistic algorithm, proposed in \cite{widmer-extreme-net}, and introduce two topology-aware heuristics to determine the number of encoded packets, that each node should forward, in order for the receivers to decode the original packets. The algorithm also allows the encoding of packets from different sources by incorporating rules for the distributed management of packet generations. Other techniques extend this algorithm by adding a feedback mechanism that enhances reliability\cite{widmer_rlnc-update} or by modifying the forwarding heuristics and the generation management mechanism\cite{mahmood_arlnccf_icc}. Another subclass of RLNC-based algorithms\cite{hou2008adapcode,cho_DRAGON,yang2011-RCODE-jrn,subramanian2012uflood} allows only the encoding of packets from the same source, which significantly limits the coding gains\cite{mahmood_arlnccf_icc}. Furthermore, algorithms in this subclass focus on reliability and integrate some kind of feedback mechanism with local scope. The implementation of such a mechanism is not straightforward in mobile networks. Therefore, those algorithms have only been proposed for static networks.

On the other hand, there are two major subclasses of XOR-based schemes. The first consists of algorithms that adopt the use of rateless codes\cite{DiSC,rahnavard2006CRBCast,vellambi2010FTS}, such as LT codes\cite{luby2002LT}.
Rateless coding requires feedback information. Therefore, similar to RLNC-based algorithms with a feedback mechanism, algorithms of this subclass have been proposed only for static networks.
The second subclass of XOR-based algorithms\cite{CodeB,Directional,kadi2008mpr-nc} follows the concept of ``coding opportunity"\cite{COPE} to perform coding on a hop-by-hop basis. The prominent algorithm of this subclass, CodeB\cite{CodeB}, combines deterministic broadcasting, based on a CDS, with hop-by-hop XOR coding of packets. It also provides information exchange mechanisms that make possible the implementation on mobile environments. The rest of the algorithms in this subclass also employ deterministic broadcasting, however they differ in the algorithm used for constructing the CDS\cite{Directional,kadi2008mpr-nc} and the rules used for finding coding opportunities\cite{kadi2008mpr-nc}. Hereafter, we use the term ``XOR-based coding" to refer to this subclass of algorithms.

\section{Analysis of RLNC's coding features}\label{motivation}
As mentioned previously, the driving force of this work has been the observation that RLNC is capable of providing robust coding features. To validate this view, we develop an analytical model that portrays the performance of RLNC in the context of broadcasting.
Before continuing with the analysis, we briefly describe random linear coding as well as the network model. Table~\ref{analnottable} summarizes the notation used in the following.
\begin{table}
\caption{Notation used in the Analysis}
\label{analnottable}
\centering
\vspace{-5pt}
\begin{tabular}{|p{0.12\columnwidth}|p{0.78\columnwidth}|}
\hline
$g$ & Generation size\\
\hline
$\mathbb{G}_{v,i}$ & Decoding matrix of node $v$ for generation $i$\\
\hline
$N$ & Number of nodes in network\\
\hline
$\mathcal{N}(v)$ &Set of node $v$'s direct neighbors\\
\hline
$\omega$ &Probability of forwarding a message\\
\hline
$\rho$ &Probability of transmission failure\\
\hline
\end{tabular}
\vspace{-12pt}
\end{table}

\subsection{Preliminaries}
\subsubsection{RLNC related concepts}

RLNC is based on the observation that a linear code, i.e. to linearly combine packets based on the theory of finite fields, is adequate for providing the benefits of network coding\cite{linear}. In order to practically implement RLNC, \emph{native}, i.e. non encoded, packets need to be organized in groups, the so called \emph{generations}\cite{PracticalNC}. Then, an encoded packet is produced as a linear combination of the native packets in a generation, using $\mathbb{F}_{2^{s}}$ arithmetic. That is, after partitioning packets into symbols of $s$ bits then $e(k)\!=\!\sum_{i=1}^{g}c_{i}p_{i}(k), \forall k$, where $e(k)$ and $p_{i}(k)$ are the $k$-th symbols of the encoded and the $i$-th native packet and $g$ is the number of packets in a generation. The set of coefficients $\langle c_{1},c_{2},\ldots,c_{g}\rangle$, which is  called \emph{the encoding vector}, is randomly selected  and appended to the packet header. The random selection provides the required flexibility for distributed implementations. It is also sufficient since the probability of producing linearly dependent packets depends on the field size $2^{s}$\cite{Ho_randomized} and is negligible even for small values of $s$\cite{Wu_tree-packing}.
Decoding packets of generation $i$ at node $v$ is performed by means of a decoding matrix $\mathbb{G}_{v,i}$, which is populated by received encoded packets. This is accomplished by performing the Gaussian elimination when $\mathbb{G}_{v,i}$ has a full rank. Decoding a subset of packets is also possible when a full rank submatrix of $\mathbb{G}_{v,i}$ exists (\emph{partial decoding}). Furthermore, encoding at an intermediate node is possible without the need of decoding the native packets since a new encoded packet may be produced by linearly combining other encoded packets.

\subsubsection{Network model}

In this work we focus on random geometric graphs (RGGs)\cite{random-penrose}. The nodes are deployed over some area $A\!\times\!A$ and an edge between a node pair $(u,v)$ exists when the Euclidean distance $d(u,v)$ is smaller than a transmission range $R$. Such a network is modelled as a graph $G(\mathcal{V},\mathcal{E})$, where $\mathcal{V}\;(|\mathcal{V}|\!\!=\!\!N)$ is the set of nodes and the set of edges is $\mathcal{E}\!\!=\!\!\{(u,v)\;|\;d(u,v)\!\leq\!R\}$. The neighborhood $\mathcal{N}(v)$ of a node $v$ is the set of nodes connected to $v$ with an edge, i.e. $\mathcal{N}(v)=\{u\;|\;(u,v) \in \mathcal{E}\}$. We focus on the generic approach of uniform node deployment which captures static and some cases of mobile networks (e.g. when node movement follows the random direction model\cite{rd-dist}). Moreover, our study is also valid for the node distribution that results from the random waypoint movement model\cite{rwp-dist}.
A node may act as a source and broadcast a message, while all other nodes implement a simple probabilistic protocol, i.e. forward the message with probability $\omega$.
The network consists of unreliable links. The transmission of a packet over a link fails with probability $\rho$, which is independent of other links.

\subsection{Performance analysis}\label{performance-analysis}
\subsubsection{Distribution of the number of message copies}

The properties of an RGG are critical for the performance of RLNC. More specifically, we will show that the performance of RLNC depends on the number of message copies that a node $d$ receives when a source $s$ broadcasts a message without using network coding. Let us model this number as a discrete RV, denoted as $X$. We aim at identifying a good approximation for the probability mass function (pmf) of $X$.
We first assume lossless links (i.e. $\rho\!\!=\!\!0$) and later generalize our model to include the case of $\rho\!\!\neq\!\!0$.
First, note that $X$ is conditional on the number of $d$'s neighbors that receive at least one copy of the broadcast message. If $Y$ is a RV representing the latter number, then $X$ follows the binomial distribution with parameters $Y$ and $\omega$, i.e. $X\!\!\sim\!\!B(Y,\omega)$. This is because the forwarding decisions made by neighbors are independent.
Then, we focus on research efforts that have established, by means of percolation theory, that probabilistic forwarding presents a bimodal behavior\cite{Gossip}. That is, if we consider the number of nodes ($r$) that receive the message, then, with high probability, either $r\!\!=\!\!0$ or $r\!=\!\alpha, \{\alpha\!\in\!\mathbb{N}\!:\!0\!<\!\alpha\!\leq\!N\}$. The probability that $r$ has any other value is negligible. The actual probability of $r\!\!=\!\!0$ (and the complementary of $r\!\!=\!\!\alpha$) as well as $\alpha$ depend on the network properties. Moreover, in most cases, $\alpha\!\rightarrow\!N$, i.e. either none or nearly all the nodes receive the message\cite{Gossip}.
By extending this finding, we make the observation that $Y$ also exhibits a near bimodal behavior, therefore a good approximation for $Y$'s pmf is:
\begin{equation}\label{intermediate_pmf}
\mathrm{P}\{Y\!=\!k\}\!=\!\begin{cases}
\phi &\!\!k\!=\!0\\
1-\phi&\!\!k\!=\!|\mathcal{N}(d)|\\
0&\!\!\textrm{otherwise}
\end{cases}
\end{equation}
where $\{\phi\!\in\!\mathbb{R}\!:\!0\!\leq\!\phi\!\leq\!1\}$. The rationale behind this approximation is simple; due to spatial proximity of the nodes that belong to $\mathcal{N}(d)$, all of them will lie either in the set of receivers or in the set of non-receivers with high probability.
Using (\ref{intermediate_pmf}), it is easy to show\footnote{Observe that $X$ can be seen as a set of $Y$ i.i.d. Bernoulli RVs. Then, $G_{X}(z)\!=\!G_{Y}(G_{B}(z))$, where $G$ denotes the probability generating function and $B$ indicates a Bernoulli RV} that:
\begin{equation}\label{first_pmf}
\mathrm{P}\{X\!=\!k\}\!=\!\begin{cases}
\phi\!+\!(1\!-\!\phi)(1\!-\!\omega)^{|\mathcal{N}(d)|}&\!\!k\!=\!0\\
(1\!-\!\phi)\omega^{k}(1\!-\!\omega)^{|\mathcal{N}(d)|-k}&\!\!0\!<\!k\!\leq \!|\mathcal{N}(d)|
\end{cases}
\end{equation}
To validate this distribution, we need to examine it under various combinations of $\omega$, $|\mathcal{N}(d)|$, the average node degree and the hop distance ($H$) between $s$ and $d$. This is because $\phi$, in analogy to the bimodal property\cite{Gossip}, also depends on those parameters. Therefore, we adopt the following strategy; we simulate probabilistic broadcasting for various values of $\omega$ in RGGs deployed in areas of various sizes (we use the normalized value $\hat{A}\!\!=\!\!A/R$ to denote the size of the network area). Then, for each $\langle \omega,\hat{A}\rangle$ pair we execute $10^{6}$ simulations. In each simulation we create a new RGG, randomly select a source-destination pair $(s,d)$ and record the number of message copies received by $d$. For each combination of $\omega,\hat{A},H,|\mathcal{N}(d)|$, we construct the statistical pmf based on the frequency observed for each value of $X$. Let $\widetilde{\mathrm{P}}\{X\!\!=\!k\}$ denote this pmf. We approximate
$\phi$ in (\ref{first_pmf}) by solving $\phi\!+\!(1\!-\!\phi)(1\!-\!\omega)^{|\mathcal{N}(d)|}\!=\!\widetilde{\mathrm{P}}\{X\!\!=\!0\}$ and calculate the total variation distance ($d_{TV}$) between (\ref{first_pmf}) and $\widetilde{\mathrm{P}}\{X\!\!=\!k\}$.
\renewcommand{\tabcolsep}{1pt}
\begin{table}
 \footnotesize
  \centering
  \caption{Total variation distance ($\scriptstyle{\times 10^{\textrm{-}\scriptscriptstyle{2}}}$) of the Approx. Distribution} \label{totaldist}
  \vspace{-8pt}
\begin{tabular}{!{\vrule width 0.5pt}c !{\vrule width 0.5pt} c|c|c|c !{\vrule width 0.5pt} c|c|c|c|c !{\vrule width 0.5pt} c|c|c|c|c!{\vrule width 0.5pt}}
  \noalign{\hrule height 0.5pt}
  & \multicolumn{4}{c !{\vrule width 0.5pt}}{$\hat{A}\!=\!4$} & \multicolumn{5}{c !{\vrule width 0.5pt}}{$\hat{A}\!=\!6$} &\multicolumn{5}{c!{\vrule width 0.5pt}}{$\hat{A}\!=\!8$}\\
  \hline
  $\omega$ & $\scriptscriptstyle{|\!\mathcal{N}\!(\!d)\!|}$ & $\scriptscriptstyle{H=2}$ & $\scriptscriptstyle{H=3}$ & $\scriptscriptstyle{H=5}$ & $\scriptscriptstyle{|\!\mathcal{N}\!(\!d)\!|}$ & $\scriptscriptstyle{H=2}$ & $\scriptscriptstyle{H=4}$ & $\scriptscriptstyle{H=6}$ & $\scriptscriptstyle{H=8}$ & $\scriptscriptstyle{|\mathcal{N}\!(\!d)\!|}$ & $\scriptscriptstyle{H=2}$ & $\scriptscriptstyle{H=4}$ & $\scriptscriptstyle{H=7}$ & $\scriptscriptstyle{H=9}$\\
  \hline
0.9&	\multirow{5}{*}{6}&	0.99&	1.21&	0.90&	\multirow{5}{*}{4}&	0.93&	0.82&	1.49&	0.37&	\multirow{5}{*}{2}&	0.77&	1.58&	2.06&	1.35\\
\hhline{-~---~----~----}
0.7&	&	0.56&	0.84&	1.32&	&	0.68&	1.00&	0.51&	1.01&	&	2.24&	2.16&	1.13&	0.43\\
\hhline{-~---~----~----}
0.5&	&	1.28&	1.61&	1.40&	&	2.74&	1.40&	0.82&	0.60&	&	1.44&	0.39&	0.03&	0.02\\
\hhline{-~---~----~----}
0.3&	&	1.08&	1.99&	1.72&	&	1.29&	0.22&	0.07&	0.01&	&	0.17&	0.06&	0.05&	0.01\\
\hline
0.9&	\multirow{5}{*}{14}&	0.42&	0.12&	0.73&	\multirow{5}{*}{8}&	0.39&	0.17&	0.13&	0.69&	\multirow{5}{*}{4}&	0.75&	0.75&	0.78&	1.46\\
\hhline{-~---~----~----}
0.7&	&	0.29&	0.45&	0.47&	&	1.02&	0.91&	0.58&	0.66&	&	2.04&	1.58&	1.01&	1.01\\
\hhline{-~---~----~----}
0.5&	&	0.62&	0.33&	0.62&	&	2.72&	1.68&	1.29&	0.73&	&	1.81&	1.12&	0.62&	0.27\\
\hhline{-~---~----~----}
0.3&	&	2.37&	1.88&	1.14&	&	3.45&	0.92&	0.20&	0.15&	&	1.01&	0.13&	0.02&	0.02\\
\hline
0.9&	\multirow{5}{*}{22}&	0.66&	0.58&	1.78&	\multirow{5}{*}{12}&	0.61&	0.95&	0.99&	1.79&	\multirow{5}{*}{7}&	1.10&	0.88&	0.45&	1.16\\
\hhline{-~---~----~----}
0.7&	&	0.64&	0.22&	1.64&	&	0.92&	0.73&	0.46&	0.78&	&	2.63&	1.41&	1.02&	0.58\\
\hhline{-~---~----~----}
0.5&	&	0.52&	0.47&	2.20&	&	1.42&	1.49&	1.58&	1.40&	&	3.57&	1.83&	0.30&	0.34\\
\hhline{-~---~----~----}
0.3&	&	2.07&	1.92&	1.37&	&	5.20&	1.25&	0.48&	0.21&	&	2.89&	0.37&	0.08&	0.02\\
\noalign{\hrule height 0.5pt}
\end{tabular}
\vspace{-12pt}
\end{table}

Table~\ref{totaldist} reports $d_{TV}$ values for networks with $\hat{A}\!\!=\!\!\{4,6,8\}$ and $N\!\!=\!\!100$. The lower value of $\hat{A}$ corresponds to relatively dense networks while the highest has been chosen so that the resulting networks are as sparse as possible but not partitioned with high probability\cite{appel-connected}. We have obtained similar results for various values of $N$, however, for brevity, we report only the results for $N\!\!=\!\!100$.
According to the presented results, (\ref{first_pmf}) provides a satisfactory approximation for the purposes of the following analysis.
As a final note, (\ref{first_pmf}) can be generalized to include the case of transmission errors, i.e. when $\rho\!\neq\!0$. Simulation results (omitted for brevity) confirm that the approximation is still good if $\omega$ is replaced by $\omega(1\!-\!\rho)$. Furthermore, we have also obtained results confirming that (\ref{first_pmf}) is still valid when nodes' positions follow the random waypoint distribution\cite{rwp-dist}. This is in accordance with a similar observation regarding the bimodal behavior of probabilistic broadcasting under the same node distribution\cite{Gossip}.

\subsubsection{Delivery Efficiency}
The performance of RLNC depends on the ability of a node to fully or partially decode a generation, which in turn depends on the rank of the decoding matrix. We examine the usual approach in the literature, in which a source node transmits a new encoded packet each time a native packet is created and added in a generation. In the context of RLNC, each intermediate node, instead of forwarding a received encoded packet, creates a new one. As a result, each node will receive a number of encoded packets with a probability given by (\ref{first_pmf}). The received encoded packets may increase the rank of the decoding matrix, depending on whether they are innovative or not. We assume that the delay from a source to a receiver is smaller than the time between the creation of two native packets, so that all the encoded packets, created after adding the $(k\!-\!1)$-th native packet, arrive before the ones created after adding the $k$-th. Then, the rank of the decoding matrix can be modeled as a stochastic process $Z\!=\!\{Z_{k},k\!\in\!\mathbb{N}\}$, where the RV $Z_{k}$ denotes the rank after a node $d$ receives all the encoded packets created by the $k$-th native packet. Note that $Z$ is memoryless because $Z_{k}$ depends only on the total number of innovative packets received after $k\!-\!1$ native packets (i.e. $Z_{k-1}$). Therefore, $Z$ is a discrete-time Markov chain and its state space is $[0,g]\!\in\!\mathbb{N}$.
\begin{figure}
  \center
  \includegraphics[width=0.99\linewidth]{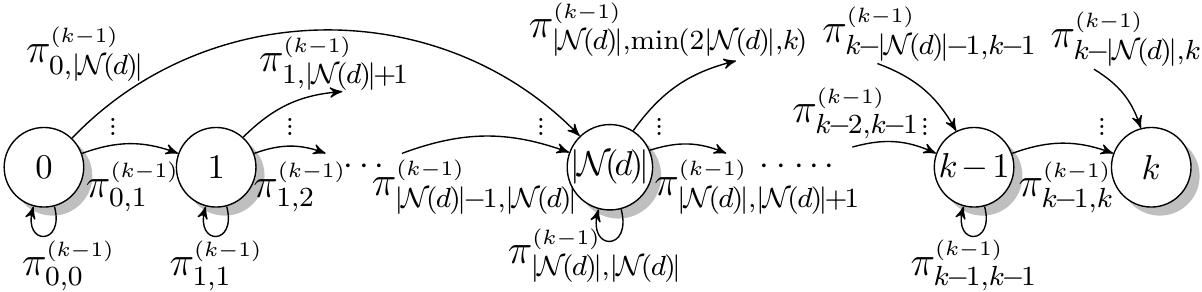}
  \caption{The proposed Markov chain at time $\big(k\!-\!1,k], k>|\mathcal{N}(d)|$.} \label{markov_chain}
  \vspace{-12pt}
\end{figure}
In the following, we focus on analysing the best case performance of RLNC in order to illustrate its full potential for providing increased delivery efficiency. We discuss the case of non-optimal performance in Section~\ref{probabilistic-harmful}.
Suppose that, at time $k\!-\!1$, the rank of the decoding matrix of node $d$ is $i$, i.e. $Z_{k-1}\!=\!i$, and that $d$ receives $j\!-\!i\leq |\mathcal{N}(d)|$ encoded packets. If $n_{in}$ denotes the number of the encoded packets that are also innovative, then $Z_{k}\!=\!i+n_{in}$. Note that $n_{in}\!\leq\!j\!-\!i$. Furthermore, $Z_{k}\!\leq\!k$ because at time $k$ only $k$ native packets have been added in the generation. This implies that $n_{in}\!\leq\!k\!-\!i$. Therefore, $n_{in}\!\leq\!\min\{j\!-\!i,k\!-\!i\}$. The best performance occurs when the rank of the matrix is maximal or, equivalently, $n_{in}$ is maximal. When $j\!<\!k$, the best performance is when $n_{in}\!=\!j\!-\!i$, i.e. all the received encoded packets are also innovative. In this case, $Z_{k}\!=\!i\!+\!(i\!-\!j)\!=\!j$ and the transition probability from state $i$ to state $j$ is therefore equal to the probability or receiving $j\!-\!i$ encoded packets. However, when $j\!\geq\!k$, only $k\!-\!i$ out of the $j\!-\!i$ encoded packets are innovative because $Z_{k}$ cannot exceed $k$. In this case, $Z_{k}\!=\!i\!+\!(k\!-\!i)\!=\!k$ and the transition probability from state $i$ to state $k$ is equal to the probability of receiving $k\!-\!i$ or more encoded packets. Summarizing, the transition probabilities in the interval $(k\!-\!1,k], \;1\!\leq\!k\!\leq\!g$ are:
\begin{equation}\label{transition_probabilities}
\pi_{i,j}^{(k-1)}\!=\!\begin{cases}
\mathrm{P}\{X\!=\!j-i\}& j\!-\!i\!\leq\!|\mathcal{N}(d)|,j\!<\!k,i\!<\!k\\
\vspace{3pt}
\sum\limits_{w=k-i}^{|\mathcal{N}\!(\!d)|}\mathrm{P}\{X\!\!=\!\!w\} & j\!-\!i\!\leq\!|\mathcal{N}(d)|,j\!=\!k,i\!<\!k\\
0&\!\!\textrm{otherwise}
\end{cases}
\end{equation}
For $k\!>\!g$, $\pi_{i,i}^{(k\mkern-1.5mu-\mkern-1.5mu1)}\!=\!1$ and $\pi_{i,j}^{(k\mkern-1.5mu-\mkern-1.5mu1)}\!=\!0, \forall j\!\neq\!i$ since after time $k\!=\!g$ no native packets are added in the generation. Note that the Markov chain is time-inhomogeneous. Fig.~\ref{markov_chain} illustrates the transition probabilities for the time interval $\big(k\!-\!1,k], k>|\mathcal{N}(d)|$.
The initial distribution is $\mathrm{P}\{Z_{0}\!=\!0\}\!=\!1$ and $\mathrm{P}\{Z_{0}\!=\!i\}=0, \forall i>0$. Therefore:
\begin{equation}\label{marginal_distribution}
  \mathrm{P}\{Z_{k}=i\}=\sum\limits_{w=0}^{g}p_{w,i}^{(k)}\mathrm{P}\{Z_{0}=w\}=p_{0,i}^{(k)}
\end{equation}
where $p_{0,i}^{(k)}$ is the element of table $\mathbf{\Pi}^{\scriptscriptstyle{(k)}}\!=\!\mathbf{\boldsymbol\pi}^{\scriptscriptstyle{(0)}}\mathbf{\boldsymbol\pi}^{\scriptscriptstyle{(\mkern-1mu1\mkern-1mu)}}\cdot\cdot\cdot\mathbf{\boldsymbol\pi}^{\scriptscriptstyle{(k\mkern-1mu-\mkern-1mu1)}}$ in the position $(0,i)$ and $\mathbf{\boldsymbol\pi}$ are the transition matrices constructed using (\ref{transition_probabilities}). Decoding is possible when $Z_{k}\!\!=\!\!k$ because $k$ innovative packets are required for decoding the $k$ native packets that exist in a generation at time $k$\footnote{We underestimate the decoding probability since decoding may be possible even if $Z_{k}\!<\!k$}. Furthermore, decoding of exactly $k$ packets occurs when $Z_{k}\!=\!k$ but no further decoding is possible, i.e. $Z_{w}\!<\!w, \forall w\!>\!k$. As a result, the expected delivery rate is:
\begin{equation}\label{expected_del}
  D_{R}=\frac{\sum\limits_{k=1}^{g}\Big[k\mathrm{P}\{Z_{k}\!=\!k\}\prod\limits_{w=k+1}^{g}(1-\mathrm{P}\{Z_{w}\!=\!w\})\Big]}{g}
\end{equation}
where $\prod_{w=k+1}^{g}(1\!\!-\!\!\mathrm{P}\{Z_{w}\!\!=\!\!w\})$ is the probability that no decoding is possible for $w\!\!>\!\!k$.

In XOR-based coding, packets are encoded under the requirement that each recipient node will decode at maximum one native packet. In other words, receiving an encoded packet is equivalent to receiving a copy of a native packet. Since receiving a single copy is enough, the expected delivery rate is $D_{X}\!=\!1\!-\!\mathrm{P}\{X\!=\!0\}$. Note that, this is the best case performance as we do not take into account decoding failures.

\subsection{Probabilistic broadcasting considered harmful}\label{probabilistic-harmful}
\begin{figure}
\centering
\subfigure[]{\includegraphics[width=0.49\linewidth]{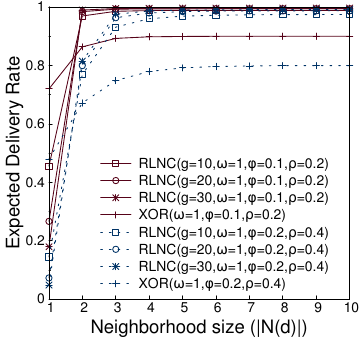}\label{theoretical_performance_flooding}}
\subfigure[]{\includegraphics[width=0.49\linewidth]{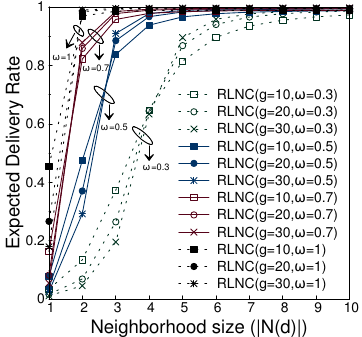}\label{theoretical_performance_gossip}}%
\caption{Analysis of RLNC's delivery efficiency: (a) comparison with XOR-based coding, (b) impact of probabilistic forwarding ($\phi\!\!=\!\!0.1$,$\rho\!\!=\!\!0.2$).}
\vspace{-12pt}
\label{theoretical_performance}
\end{figure}
Fig.~\ref{theoretical_performance_flooding} illustrates the expected delivery rate for RLNC and XOR-based coding when combined with flooding ($\omega\!=\!1$). More specifically, the delivery rate is plotted versus the node degree using different values of $\phi$ and $\rho$. RLNC exhibits high levels of resilience to transmission impairments and dominates XOR-based coding even when $\phi$ and $\rho$ increase. A simple explanation is that using RLNC in broadcasting enables a node to exploit message redundancy (or equivalently path diversity) to recover not just a single packet but any packet from the generation. This is possible through the creation of new encoded packets in each intermediate node, which allows a node to receive a plethora of possibly useful packets. This also explains why RLNC fails when message redundancy is absent ($|\mathcal{N}(d)|\!=\!1$). RLNC-based schemes should specially treat such cases. As a final note, the proposed Markov chain can be easily generalized to describe the cases that a node receives less than the maximum number of innovative packets per native one. Our results indicate that RLNC continues to dominate the best-case performance of XOR-based coding (for a wide range of $\phi$ and $\rho$ values). The only exception is when a node receives, at maximum, only one innovative packet for each native, i.e. again when diversity is absent. Nevertheless, such a situation is highly unlikely.

The advantage of XOR-based schemes is the reduced cost, since coding is utilized towards reducing transmissions. Instead, RLNC-based schemes resort to probabilistic forwarding for reducing cost. Fig.~\ref{theoretical_performance_gossip} depicts the performance of RLNC when combined with probabilistic forwarding for different values of $\omega$. The plotted results are in accordance with simulation data reported in \cite{widmer-extreme-net}. Clearly, pruning transmissions significantly impairs RLNC's performance. This performance degradation has also been identified, implicitly\cite{fragouli_rlnc} or explicitly\cite{widmer_rlnc-update}, however the problem has been treated within the context of probabilistic broadcasting. We believe that the key factor for RLNC's performance degradation is the unsystematic way of pruning transmissions, which does not take into account information about connectivity. The strategy to prune transmissions based on heuristics that account for the node degree\cite{fragouli_rlnc} is towards the correct direction. However, we feel that such heuristics should also take into account topology-related information of non-local scope. Some of this information is difficult to obtain and even if this was possible, it would require a complex analytical model to define the optimal $\omega$. Therefore, \emph{we opt for a more systematic and self-configuring pruning mechanism that takes into account the network topology}.

\section{The Synergy of RLNC and Deterministic Broadcasting}\label{RLDP}
Following the previous observations, we adopt RLNC. Yet, contrary to the common approach, we implement it on top of a deterministic broadcast algorithm. We choose Partial Dominant Pruning (PDP)\cite{PDP} from the class of Dominant Pruning (DP) algorithms. DP algorithms distributively construct a CDS in order to broadcast messages. Our intuition is that the CDS will provide a topology-aware, self-configuring process for reducing transmissions. However, establishing this synergy, without damaging RLNC's coding efficiency, is not a trivial task. PDP's forwarding rules need to be redesigned so as to treat packets as members of a group, i.e. the generation. \emph{Random Linear network coding over Dominant Pruning (RLDP)} incarnates the aforementioned concepts.

\subsection{Dominant Pruning fundamentals}
In DP algorithms, a node $v$, with a message to broadcast, decides which of its neighbors should act as forwarders and informs them by piggybacking on the message the corresponding list, called the \emph{forwarding set} ($fs(v)$). The process is then repeated by every forwarder until a \emph{termination criterion} is met\cite{PDP}. Forwarders should be elected so as to deliver the message to (or ``cover" according to the set cover terminology) the set of nodes that lie exactly 2-hops away from $v$. This latter set is also called the universal set $\mathcal{U}(v)$, i.e. $\mathcal{U}(v)\!\!=\!\!\mathcal{N}(\mathcal{N}(v))\!-\!\mathcal{N}(v)$, where $\mathcal{N}(\mathcal{N}(v))$ is the set of nodes lying within 2-hops from $v$. The set of candidate forwarders $\mathcal{C}(v)$ consists of $v$'s neighbors, i.e. $\mathcal{C}(v)\!\!=\!\!\mathcal{N}(v)$. Note that $\mathcal{U}(v)\!\!\subseteq\!\bigcup_{\forall u\in \mathcal{C}(v)}(\mathcal{N}(u)\!-\!\mathcal{N}(v))$ and that $\mathcal{C}(v)$ can be seen as a set of sets if each node $u\in \mathcal{C}(v)$ is replaced by $\mathcal{N}(u)\!-\!\mathcal{N}(v)$, thus the set cover problem. The problem is solved using the well-known greedy set cover (GSC) algorithm\cite{intro-algo-cormen}, however other approximation algorithms exist\cite{MPR,epap-efcn}. PDP makes the observation that, when $v$ receives a message from $u$, both $\mathcal{C}(v)$ and $\mathcal{U}(v)$ can be reduced by eliminating the nodes covered by $u$, i.e. $\mathcal{C}(v)\!\!=\!\!\mathcal{N}(v)\!-\!\mathcal{N}(u)$ and $\mathcal{U}(v)\!\!=\!\!\mathcal{N}(\mathcal{N}(v))\!\!-\!\!\mathcal{N}(v)\!\!-\!\!\mathcal{N}(u)\!\!-\!\!\mathcal{N}(\mathcal{N}(u)\cap \mathcal{N}(v))$.

\subsection{Coding Rules}
As mentioned previously, RLDP utilizes the basic functionality of random linear coding. Below, we discuss some important design choices and the rationale behind them.\\
\noindent\textit{Strictly Inter-source coding}: RLDP adopts \emph{inter-source coding} (i.e. packets from different sources are mixed in a generation) in the light of empirical evidence which prove that it increases the coding efficiency when compared to \emph{intra-source coding} (i.e. a generation contains packets from the same source)\cite{mahmood_arlnccf_icc}. Despite the advantages of inter-source coding, only a minority of algorithms\cite{widmer-extreme-net,fragouli_rlnc,mahmood_arlnccf_icc} follows this approach. The reason is that, in this case, the problem of generation management is not trivial. In contrast to the usual approach, which is to operate inter-source and intra-source coding in parallel, in RLDP, each source can add only one packet in each generation. In other words, we adopt inter-source coding but do not allow intra-source coding. We call this strategy \emph{strictly inter-source coding (SIS).} Our approach stems from the belief that inter-source coding performs better than intra-source for poorly connected source-receiver pairs. An intuitive explanation is that inter-source coding achieves a better spreading of information because it involves more sources (possibly from different parts of the networks). As a result, information about native packets is easier to bypass low connectivity areas. Section~\ref{evaluation} confirms the effectiveness of SIS.\\
\noindent\textit{Encoding and decoding}: A source node creates and transmits a new encoded packet immediately after adding a new native packet into a generation. The rationale of this strategy is twofold; first it aims to ensure that the new information carried by the native packet will be propagated through the network with minimum delay. Second, it facilitates partial decoding, which reduces end-to-end delay. To understand this, bear in mind that non zero rows of a decoding matrix correspond to innovative packets while non zero columns correspond to native packets. Consequently, the strategy of immediately transmitting a new encoded packet, increases the probability that the decoding matrix contains a full rank square submatrix, thus enabling partial decoding.

\subsection{Forwarding Rules}
\subsubsection{Propagating generations over the CDS}
In order to achieve the synergy of RLNC and PDP, we need to enable the \emph{propagation of generations through the CDS} formed by PDP. Note that, in DP algorithms, a node $v$ reacts to the reception of a packet only if it has been selected as a forwarding node. Furthermore, recall that, in RLNC, only a subset of the encoded packets, the innovative ones, carry useful information about a generation. Therefore, the first intuitive approach is to adopt the following forwarding strategy:
\newtheorem{definition}{Definition}
\begin{definition}[Innovative-based criterion]
A forwarding node produces and transmits a new encoded packet iff it receives an innovative packet.
\end{definition}
In the context of DP, the Innovative-based criterion is actually a termination criterion, i.e. the execution of the algorithm stops when a non innovative packet is received. This criterion is the analogous of the stopping conditions adopted by schemes that implement RLNC on top of probabilistic broadcasting\cite{fragouli_rlnc,mahmood_arlnccf_icc}. Given the Innovative-based criterion, we can prove the correctness of RLDP\footnote{We assume that the probability of producing linearly dependent encoded packets is negligible\cite{Wu_tree-packing}. This assumption is common in the related literature.}, i.e. that all network nodes can fully decode a generation in a lossless network. First, we prove that:
\newtheorem{lemma}{Lemma}
\begin{lemma}\label{lemma1}
Every node $v$, which is not the source node of a native packet $q$, receives \emph{at least} one innovative packet after $q$ is added in a generation.
\end{lemma}
\begin{IEEEproof}
The source node $s$, after adding $q$ to a generation $i$, defines a forwarding set $fs(s)\!\!=\!\!\{f_{1}, f_{2},\ldots\}$ and transmits a new encoded packet $e_{s,i}$. Every node $v\!\in\!\mathcal{N}(s)$ will receive this packet, which is innovative since it ``contains" $q$. Furthermore, the solution of the set cover problem guarantees that, given a node $u\!\!\in\!\!\mathcal{N}(\mathcal{N}(s))$, there is at least one forwarder $f\!\in\!fs(s)$ that covers $u$. Since $f\!\!\in\!\!\mathcal{N}(s)$, it will receive the innovative packet $e_{s,i}$ and will transmit a new encoded packet $e_{f,i}$. As a result, each node $u\in \mathcal{N}(\mathcal{N}(s))$ will receive at least one encoded packet $e_{f,i}$. The first of these packets is clearly innovative since it ``contains" $q$. The same reasoning can be used in subsequent hops to include all network nodes.
\end{IEEEproof}
Moreover, we can prove that:
\begin{lemma}\label{lemma2}
Every node $v$, which is not the source node of a native packet $q$, receives \emph{exactly} one innovative packet after $q$ is added in a generation.
\end{lemma}
\begin{IEEEproof}
According to Lemma~\ref{lemma1}, if $g$ native packets are added in generation $i$, then each node $v$ will receive $g'\!\!\geq\!\!g$ innovative packets. It suffices to show that $g'\!\!=\!\!g$. Note that the row rank of $\mathbb{G}_{v,i}$ (which equals $g'$) cannot exceed the column rank (which equals $g$), i.e. $g'\!\!>\!\!g$ is not possible.
\end{IEEEproof}
We use this lemma to prove that:
\newtheorem{theorem}{Theorem}
\begin{theorem}[Correctness of RLDP]\label{theorem_correctness}
Every node can decode a generation in a lossless network.
\end{theorem}
\begin{IEEEproof}
Lemma~\ref{lemma2} secures that, any non source node $v$ will receive exactly $g$ innovative packets for a generation of size $g$, thus $\mathbb{G}_{v,i}$ has a full rank. Furthermore, each source $s$ will receive exactly $g\!\!-\!\!1$ innovative packets, one for each native packet added by the other sources. This is sufficient since $s$ only needs to decode $g\!\!-\!\!1$ packets.
\end{IEEEproof}

\subsubsection{Reducing transmissions}
\begin{figure}
  \center
  \includegraphics[width=0.7\linewidth]{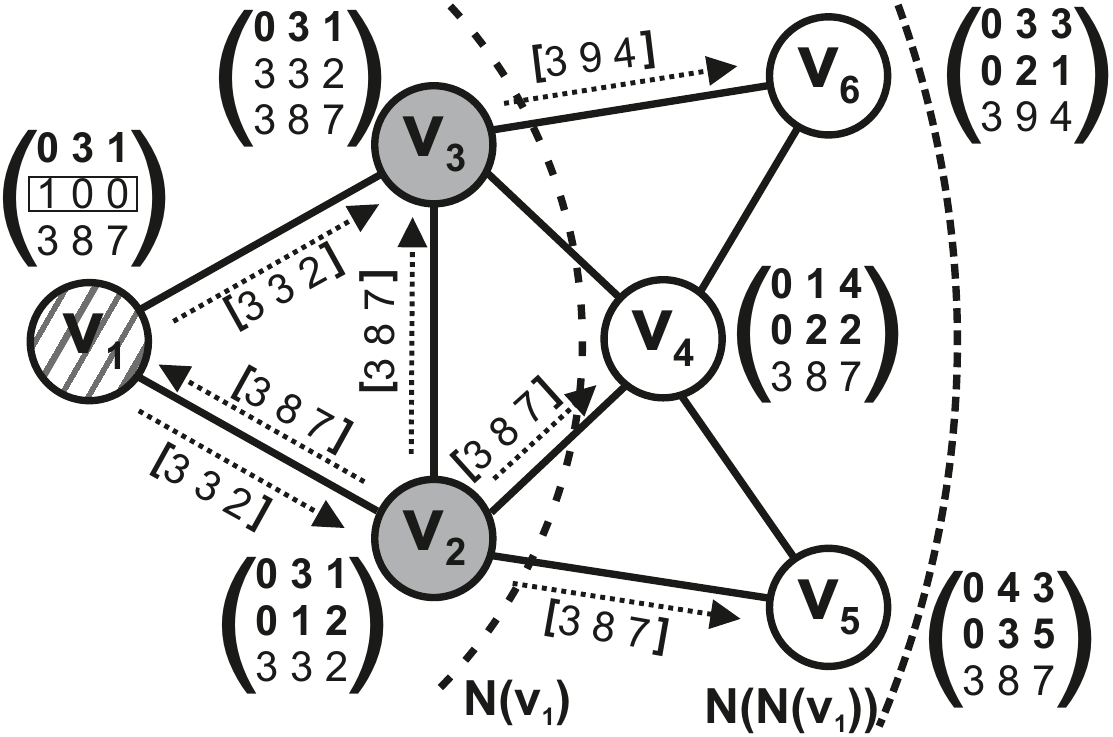}
  \caption{Example of broadcasting with RLDP (Single-Innovative criterion)}\label{forwarding_intuition_example}
  \vspace{-12pt}
\end{figure}

According to Lemma~\ref{lemma2}, the Innovative-based criterion is equivalent to the strategy of forwarding one encoded packet for each native one added in a generation. However, in the presence of transmission errors, if a native packet is added in a generation, a node $v$ will receive more than one innovative packet. This happens when the rank of its decoding matrix is lower than the rank of the decoding matrices of its neighbors. Using the Innovative-based criterion in such cases will result in $v$ transmitting more than one encoded packet for each native one. To explain the situation, let us examine the example in Fig.~\ref{forwarding_intuition_example}. In this example, $g\!=\!3$ and the bold rows in the decoding matrices indicate the encoding vectors of the innovative packets that have already been received. Note that, due to transmission errors, $v_{1}$ and $v_{3}$ have received only one innovative packet. At some point, $v_{1}$ acts as a source, adds a packet in the generation and after selecting the forwarding set (in this case $fs(v_{1})\!\!=\!\!\{v_{2},v_{3}\}$), transmits an encoded packet (the innovative packets that are transmitted in the network are illustrated with dashed lines along with the corresponding encoding vectors). Note that $v_{3}$ will receive two innovative packets (one from $v_{1}$ and one from $v_{2}$) and, according to the Innovative-based criterion, should transmit two encoded packets. We make the observation that, in the context of Dominant Pruning, \emph{not all innovative packets need to result in the transmission of a new encoded packet}. In fact, we introduce
the following policy:
\begin{definition}[Single-innovative criterion]
A forwarding node produces and transmits a new encoded packet only for the first innovative packet that is received as a result of the addition of a native packet in a generation.
\end{definition}
The rationale of this policy is clear and, in part, is expressed by Lemma~\ref{lemma2} and Theorem~\ref{theorem_correctness}; in the absence of transmission errors only one innovative packet per native is adequate while, in the presence of transmission errors, \emph{a node should only rely on the path diversity provided by the network to recover from transmission errors}. To further explain, let us go back to the example of Fig.~\ref{forwarding_intuition_example}. When the Single-innovative criterion is used, $v_{3}$ receives two innovative packets and decodes the generation. However, $v_{3}$ will transmit only one new encoded packet. Note that this new packet is enough for $v_{6}$ to decode the generation. Furthermore, observe that $v_{6}$ actually receives two encoded packets (the second one is from $v_{4}$ and is not illustrated since it is not innovative). If the rank of $v_{6}$'s decoding matrix was initially one, both of the received packets would be innovative. Therefore, $v_{6}$ could take advantage of path diversity and decode the generation. Clearly, there is still the probability that a node will not be able to decode a generation. In general, this probability increases for nodes with low connectivity. One solution to eliminate failures would be to allow a node to relax the Single-innovative criterion based on the connectivity of its neighbors or even based on feedback information. However, such approaches come at the cost of extra transmissions. We refrain from investigating the impact of such heuristics, as well as the related cost, since our primary objective is to illustrate that using deterministic broadcasting, even without such heuristics, results in less decoding failures compared to a probabilistic scheme.
\begin{figure}
\begin{center}
\line(1,0){250}
\end{center}
{
\footnotesize
\vspace{-8pt}
$\;$\textbf{RLDP (prev\_node $u$, cur\_node $v$, packet $p$, generation\_id $gid$)}
\vspace{-12pt}
}
\begin{center}
\line(1,0){250}
\end{center}
\vspace{-8pt}
\begin{algorithmic}[1]
\footnotesize
\IF{($!$Innovative($p$))}
  \STATE DropPacket($p$)
\ENDIF
\STATE UpdateDecodingMatrix($p$)
\IF{($ v  \notin  p.forwarders $ $ \mid \mid $ $!$Single-innovative($p.src,gid$)}
  \STATE DropPacket($p$)
\ENDIF
\STATE $newp$=RandomLinearCoding($gid$)
\STATE $fwset$=GSC($N(v)$, $N(N(v))$, $u$)
\STATE $newp$.set($fwset$)
\vspace{-2pt}
\STATE transmit $newp $
\normalsize
\end{algorithmic}
\vspace{-12pt}
\begin{center}
\line(1,0){250}
\end{center}
\vspace{-10pt}
  \caption{Pseudocode of RLDP's forwarding procedure}\label{RLDP_pseudocode}
  \vspace{-12pt}
\end{figure}

An important issue is how to implement the Single-innovative criterion. To do so, we need to provide some kind of association between a native packet $q$ and the innovative ones produced after node $s$ adds $q$ in a generation $i$. Since in RLDP a node adds only one native packet into a generation, this task can be tackled by using the value pair $\langle s,i\rangle$, i.e. the source address and the generation id, which is contained in a packet's header. Another requirement is to allow a forwarding node $v$ to track whether an innovative packet with the same value pair $\langle s,i\rangle$ has already been received. The most efficient way is to use direct addressing\cite{intro-algo-cormen} due to the fast dictionary operations. The space complexity of such an approach ($\mathcal{O}(g)$ for a generation of size $g$) is reasonable since the generation size is usually kept low in order to reduce the decoding cost. Fig.~\ref{RLDP_pseudocode} presents the pseudocode of RLDP's forwarding procedure.

\section{Distributed Generation Management}\label{generation_management}
In order to practically implement RLNC, packets should be grouped into generations. This task, known as \emph{generation management}, is trivial when intra-source coding is used since the required decisions involve a single node and thus are made locally. However, generation management becomes more complicated in the case of inter-source coding since the sources should agree on a common grouping of packets. In general, generation management involves the following tasks:\\
\noindent\textit{1) Decide which generation to choose for adding a native packet and when to start a new generation}: Choosing a generation is the first important decision to make because it affects the overall performance. Recall that each generation is identified by an id, which is carried in each encoded packet. A generation is considered known to a node $v$ if $v$ either created the generation or has received at least one encoded packet from this generation. Let $\mathcal{GS}_{v}$ denote the set of generations which are known to $v$ and their size have not exceeded the maximum size $g$. The common approach is that a source $s$ will add a new packet to a randomly chosen generation from $\mathcal{GS}_{s}$\cite{widmer-extreme-net,fragouli_rlnc}. Another approach is to choose from a subset of $\mathcal{GS}_{s}$ which contains generations initiated from nodes that lie certain hops away from $s$\cite{fragouli_rlnc,mahmood_arlnccf_icc}. A new generation is started if $\mathcal{GS}_{v}\!\!=\!\!\emptyset$\cite{widmer-extreme-net,fragouli_rlnc,mahmood_arlnccf_icc} or when the chosen generation already contains a packet from $s$\cite{widmer-extreme-net}. All the proposed strategies aim at reaching the maximum generation size, in order to increase performance\cite{widmer-extreme-net}. However, under transmission errors, a large generation size increases the decoding delay. The reason is that it takes longer to collect the number of encoded packets that is required for decoding. Following this observation, we opt for reduced delay. Therefore, in RLDP, a source $s$ adds a new packet to the most recently seen generation, if this belongs to $\mathcal{GS}_{s}$. If no such generation is found or the selected generation already contains a packet from $s$ (strictly inter-source coding), then a new generation is created. Note that, the size of the produced generations will not necessarily be close to $g$. However, we believe that this will not have a significant impact on the decoding efficiency. This intuition is based on reported empirical data\cite{widmer-extreme-net,fragouli_rlnc}, also confirmed by the analysis in Section III, which indicate that a relatively small generation size is enough for providing the coding benefits. We confirm our intuition through simulation in Section~\ref{evaluation}.\\
\noindent\textit{2) Provide an addressing scheme for packets within a generation}: Another problem, although rarely discussed in the literature, is to uniquely identify packets within a generation. To understand this requirement, recall that each encoded packet carries an encoding vector, i.e. the coefficients $\langle c_{1},\ldots,c_{g}\rangle$ used to mix the native packets. In order for decoding to be possible, it is necessary that all nodes will be able to arrange the coefficients in the same order.
One way to secure this, is by predefining the order of coefficients. However, this is virtually impossible in a distributed environment. A more practical solution is to provide a unique id for each native packet, so as to enable sorting based on this id, and associate it with the corresponding coefficient. The simplest way to accomplish this is by using the pair $\langle node\_id,seq\_num\rangle$, where $seq\_num$ is a sequence number generated locally at the source and $node\_id$ is the source address\cite{mahmood_arlnccf_icc}. The use of $seq\_num$ enables two packets from the same source to coexist in a generation. RLDP takes a simpler approach. Since strictly inter-source coding is used, there is no way that two native packets from the same source will reside in the same generation. Therefore, only $node\_id$ can be used for uniquely identifying a packet in the generation. Our strategy, besides using a smaller identifier for packets, does not involve any overhead for managing sequence numbers.\\
\begin{figure*}[t]
\centering
\subfigure[]{\includegraphics[width=0.25\linewidth]{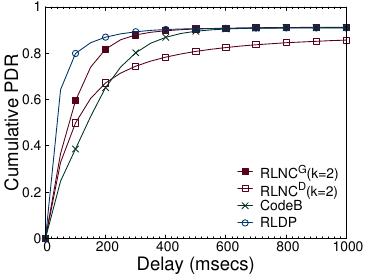}\label{CDFSparse100}}\hspace{-5pt}
\subfigure[]{\includegraphics[width=0.25\linewidth]{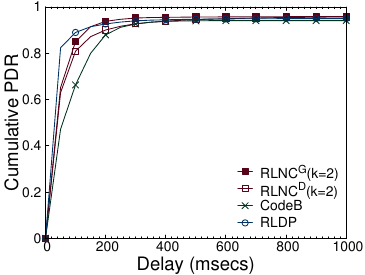}\label{CDFDense100}}\hspace{-5pt}
\subfigure[]{\includegraphics[width=0.25\linewidth]{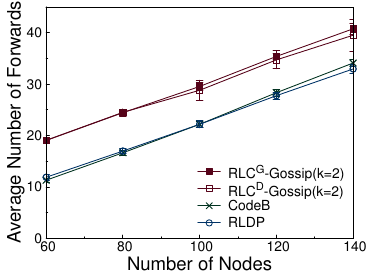}\label{FW_sparse}}\hspace{-5pt}
\subfigure[]{\includegraphics[width=0.25\linewidth]{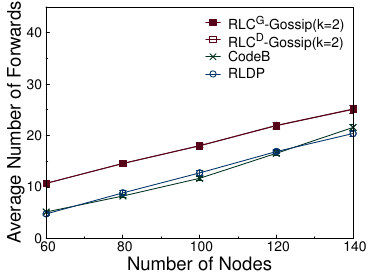}\label{FW_dense}}%
\caption{Performance for different network densities ($\lambda\!\!=\!\!1$ pkt/sec/source, max speed:1 m/sec): (a) Cumulative PDR vs delay (``Sparse",\hspace{1pt}$N\!\!=\!\!100$) (b) Cumulative PDR vs delay (``Dense",\hspace{2pt}$N\!\!=\!\!100$) (c) Avg. number of forwards vs number of nodes (``Sparse") (d) Avg. number of forwards vs number of nodes (``Dense").}
\vspace{-12pt}
\label{Topology}
\end{figure*}
\noindent\textit{3) Provide an addressing scheme for generations}: The next important problem is to uniquely identify generations so that each node can decide to which generation an encoded packet belongs to. The usual approach is that a node will randomly choose the generation id\cite{widmer-extreme-net,fragouli_rlnc,mahmood_arlnccf_icc}. Choosing from a sufficient large space minimizes the probability that two different nodes will choose the same generation id. On the contrary, in RLDP, a source does not need to randomly select the generation id. In fact, when a source starts a new generation, increases by one the most recently seen generation id and uses this as the new id. This strategy allows sources to coordinate their views about the generations in use, thus enabling them to effectively populate the generations with packets. Eliminating the random selection of generation ids allows the use of a smaller address space.

\section{Evaluation}\label{evaluation}
To evaluate RLDP's performance, we compare it with two algorithms. The first one, proposed in \cite{fragouli_rlnc}, is the most representative of RLNC-based algorithms. In the following, we will use the term RLNC to refer to this algorithm. The second algorithm, CodeB\cite{CodeB}, utilizes XOR-based coding. Regarding RLNC, we use two variants, namely RLNC$^D$ and RLNC$^G$. The first, uses the distributed generation management described in \cite{fragouli_rlnc}. In the second, we assume that each node has global coding information, i.e. perfect knowledge of the coding status of other nodes. This scheme achieves the optimal allocation of packets across generations. Although it is unrealistic, we use it to illustrate the performance bounds of RLNC. Furthermore, RLNC employs the forwarding heuristic described in \cite[Algorithm 6B]{fragouli_rlnc} with $k\!\!=\!\!2$. We chose this setting after extensive experimentation, which showed that it yields the best performance.\\
\textit{Set up and methodology}: All investigated algorithms are implemented in the ns2 simulator~\cite{ns2}, using the CMU extension. Furthermore, RLNC and RLDP were implemented based on the network coding ns2 module~\cite{ns2code}. We present the  average values over $20$ independent simulation runs, each with a duration of $900$ seconds. The confidence level, for all reported confidence intervals, is $95\%$.\\
\textit{Network model}:
The default number of nodes is $100$, the transmission range is $250$m and the nominal bit rate is $2$Mbps. The nodes move in a square area according to the Random Waypoint (RW) model~\cite{rwp-dist}. To avoid transient artifacts in nodes' movement, we use the perfect simulation algorithm~\cite{prf-sim}. We examine two network densities; ``Dense" and ``Sparse". Similar to \cite{CodeB}, in the ``Dense" topology, the average neighborhood size is 30 while in the ``Sparse" topology it is 15. Note that, we could not use a lower density in the ``Sparse" scenario since, in such a case, frequent partitions would occur. Simulations showed that in the `Sparse" scenario, there exist many nodes (those moving near the boundaries) that experience very low connectivity. All algorithms collect neighborhood information by periodically exchanging hello messages with an interval of 1 second.\\
\textit{Network traffic}: Traffic is generated by broadcast sessions, each stemming from a different source node and starting at a random time.
The size of each message is set to $256$ bytes. Furthermore, both the number of sources and the maximum generation size are fixed to 30. We chose the generation size after extensive experimentation, which showed that using a larger size does not improve performance but rather increases the related costs. We used a GF of size $2^{8}$.

\begin{figure}
\centering
\subfigure[]{\includegraphics[width=0.5\linewidth]{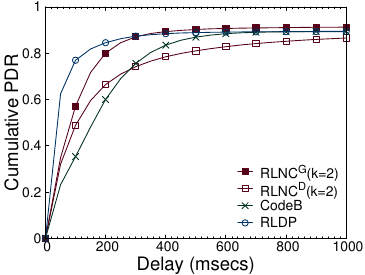}\label{CDFMob2_10}}\hspace{-5pt}
\subfigure[]{\includegraphics[width=0.5\linewidth]{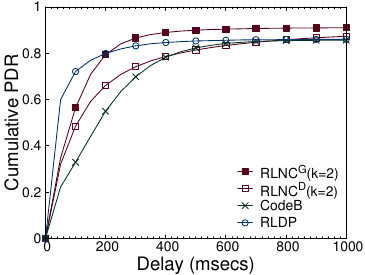}\label{CDFMob10_20}}%
\caption{Cumulative PDR vs delay ($\lambda\!\!=\!\!1$\hspace{2pt}pkt/sec/source, ``Sparse" topology). Node speed: (a) $2\!-\!10$ m/sec (b) $10\!-\!20$ m/sec.}
\vspace{-12pt}
\label{Mobility}
\end{figure}
Fig.~\ref{CDFSparse100} and \ref{CDFDense100} depict the cumulative packet delivery ratio (PDR) versus the end-to-end delay, i.e. the fraction of packets delivered within a delay limit, for ``Sparse" and ``Dense" networks. We choose this presentation style in order to capture both the delivery efficiency and the timeliness of each algorithm. The results provide a confirmation of the effectiveness of random linear network coding. Both RLDP and RLNC$^G$ outperform CodeB. The main reason is that XOR-based coding schemes introduce delay in order to detect coding opportunities. As expected, in the ``Sparse" topology, the performance of all schemes degrades. For CodeB, a low density topology reduces the coding opportunities. As a result, more transmissions occur (compare Fig.~\ref{FW_sparse} and \ref{FW_dense}) and increase the probability of collisions. In the case of random linear coding, the witnessed degradation is in accordance to the analysis in Section~\ref{performance-analysis} because in low density topologies the average neighborhood size is smaller. Nonetheless, RLDP outperforms both RLNC$^D$ and RLNC$^G$, which uses global knowledge. This justifies our approach to combine random linear coding with deterministic broadcasting. Note that, in sparse topologies, RLNC$^D$ fails to keep up with other schemes. This highlights the importance of distributed generation management. Also, observe that, RLDP's generation management does not compromise the coding gains. We tested networks of various sizes (from 60 to 140 nodes) and found qualitatively similar results. Fig.~\ref{FW_sparse} and \ref{FW_dense} illustrate the average number of forwards versus the network size for ``Sparse" and ``Dense" networks. The results confirm the intuition that the CDS, used by RLDP to forward messages, provides an efficient pruning process. More specifically, RLDP manages a reduction from $17\%$ to $38\%$ in ``Sparse" and from $19\%$ to $56\%$ in ``Dense" networks, compared to RLNC variants. Interestingly, RLDP performs similar to CodeB, despite the fact that the latter uses coding for reducing transmissions.

In the following experiments, we only examine the delivery efficiency since we observed similar findings as far as the number of forwards is concerned. Furthermore, we focus on the more challenging scenario of ``Sparse" networks. Fig.~\ref{Mobility} presents the delivery efficiency under different levels of mobility. Clearly, increased mobility levels impacts the performance of RLDP and CodeB. The reason is that both schemes use deterministic broadcasting, which is affected by topology variations. Moreover, mobility also increases the decoding failures in CodeB since successful decoding depends on the accuracy of information about the neighbors' coding status. On the contrary, RLDP minimizes the impact of mobility on the deterministic broadcasting algorithm due to the use of random linear coding. Both RLNC$^D$ and RLNC$^G$ are virtually unaffected by mobility. This is attributed to the higher message redundancy produced by the probabilistic forwarding scheme. Nevertheless, message redundancy results in a significantly increased cost (more than $30\%$ compared to RLDP) in terms of transmissions. In any case, observe that only the unrealistic RLNC$^G$ outperforms RLDP when mobility is very high.
\begin{figure}
\centering
\subfigure[]{\includegraphics[width=0.5\linewidth]{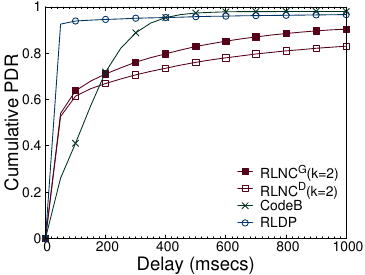}\label{CDFRate0.1}}\hspace{-5pt}
\subfigure[]{\includegraphics[width=0.5\linewidth]{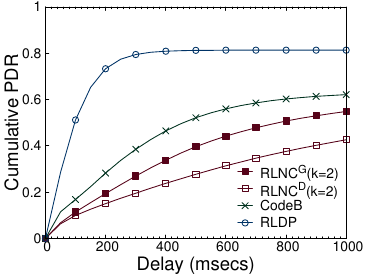}\label{CDFRate2.0}}\hspace{-5pt}
\caption{Cumulative PDR vs delay (max speed:1 m/sec, ``Sparse" topology): (a) $\lambda\!=\!0.1$ (b) $\lambda\!=\!2$ pkts/sec/source.}
\vspace{-12pt}
\label{Rate}
\end{figure}

In Fig.~\ref{Rate}, we evaluate the algorithms for different levels of traffic load. Under low traffic (Fig.~\ref{CDFRate0.1}), RLDP outperforms all algorithms. CodeB needs to wait an increased amount of time in order to find coding opportunities since fewer packets coincide in the network. Both RLNC variants suffer from increased delay as they need more time to fill the generations. On the other hand, RLDP outperforms all schemes because its generation management is oriented towards reducing delay. The tradeoff is a reduced number of packets allocated to each generation (refer to Section~\ref{generation_management}). However, this does not impair the delivery efficiency. When congestion levels increase (Fig.~\ref{CDFRate2.0}), the performance of all algorithms degrades. However, RLDP exhibits a remarkable resilience due to the combination of deterministic broadcasting and random linear coding. The former reduces the levels of congestion and thus decreases the probability of collisions. The latter uses path diversity to enhance delivery efficiency. Both mechanisms are equally important. CodeB and RLNC variants fail because they use only one of them.

\section{Conclusion}\label{conclusion}
Random linear network coding is used to enhance the resilience of protocols to packet losses. We proved, through analysis, that we need utilize topology-aware algorithm in order to maximize the benefits of random linear coding. To this end, despite the common approach in the literature, which is to use random linear coding on top of probabilistic forwarding schemes, we chose the synergy with a CDS-based broadcast algorithm. We demonstrated, through simulation, the efficiency of this approach. Moreover, we provided a distributed mechanism for managing generations. The mechanism does not compromise the coding efficiency even in networks of high mobility. In the future, we plan to investigate methods for further improving the topology-awareness of the underlying broadcasting scheme. Furthermore, we are interested in employing new generation management techniques and exploring their impact on the performance of random linear coding.

\bibliographystyle{IEEEtran}
\bibliography{IEEEabrv,rldp}

\end{document}